\documentclass[epj]{svjour}
\usepackage{graphics}
\usepackage{hyperref}
\hypersetup{
    colorlinks=true,
    urlcolor=blue,
    citecolor=black,
    linkcolor=black
}
\usepackage{csvsimple}
\usepackage{listings}
\usepackage{xcolor}
\usepackage{hyperref}
\usepackage{multirow}
\begin{document}
\newcommand{\prd} {Phys. Rev. D }
\newcommand{\apj} {ApJ }
\newcommand{\aap} {A \& A}
\newcommand{\jcap} {JCAP}
\newcommand{\mnras} {Mon. Not. Roy. Ast. Soc. }
\newcommand{\apss} {Astrophys. and Space Sci. }
\newcommand{\pasp} {Publ. of Astron. Soc. Pacific }

\lstset{language=R,breaklines=true}
\title{Median statistics estimates  of  Hubble and Newton's Constant}
\author{Suryarao Bethapudi\inst{1} \and Shantanu Desai\inst{1} 
}                     
%
%
\institute{$^{1}$Department of Physics, IIT Hyderabad, Kandi, Telangana-502285, India}
\date{Received: date}
%
\abstract{
Robustness of any statistics depends upon the number of assumptions it makes about the measured data.
We point out  the advantages of  median statistics using toy numerical experiments and demonstrate its robustness,  when the number of assumptions we can make about the data are limited.  We then apply the median statistics technique to obtain estimates of two constants of nature, Hubble Constant ($H_0$) and Newton's Gravitational Constant($G$), both of which show significant differences between different measurements.
For $H_0$, we update the analysis  done by Chen and Ratra (2011) and Gott et al. (2001) using $576$  measurements. We find after grouping the different results according to their primary type of measurement,  the median estimates are given by  $H_0=72.5^{+2.5}_{-8}$ km/sec/Mpc with errors corresponding to 95\% c.l. (2$\sigma$)  and $G=6.674702^{+0.0014}_{-0.0009} \times 10^{-11} \mathrm{N m^{2}kg^{-2}}$ corresponding to 68\% c.l. (1$\sigma$).
}
%
\maketitle
\section{Introduction}

Rapid advances in observational cosmology due to avalanche of new data have led to the era of ``precision
cosmology'' with many of the key cosmological parameters determined to about 1\% precision. The current best cosmological constraints come from Cosmic Microwave Background anisotropy measurements from the Planck
satellite~\cite{Planck}. These constraints are expected to be measured to even better precision with ongoing CMB experiments and  a variety  of optical photometric and spectroscopic surveys such as eBOSS, DES, Euclid, HSC, KiDS etc.

However, despite this, there is still no consensus for over half a century on the measurements of first ever cosmological parameter introduced in literature, viz. the Hubble constant ($H_0$), which measures the expansion rate of the universe. Until the 1990s, $H_0$ ranged from 50 to 100 km/sec Mpc. (See for example the contradictory points of views as of 1996 on measurements of $H_0$ between Tammann~\cite{Tammann} and Van Den Bergh~\cite{vandenberg}.) The tension between different measurements of $H_0$ continues to persist in 2016 in the era of precision  cosmology~\cite{Verde,Silk}. Currently there is a 3.3$\sigma$ tension between
the latest Planck constraint~\cite{PlanckH0} ($H_0=66.93 \pm 0.62$ km/s/Mpc) and local measurements from Hubble Space Telescope ($H_0=73.24 \pm 1.74$ km/s/Mpc) based on Cepheid variables~\cite{Riess}.

Similar conflicting measurements have also been reported for measurements of Newton's Gravitational Constant $G$. This constant appears in  Newton's law of universal gravitation and Einstein's field equations in General Relativity. $G$ is notoriously hard to measure and the statistical significance of the tensions between different measurements  is about $10\sigma$~\cite{Quinn,Mohr}. There have also been claims of periodicities in measurements of $G$~\cite{Anderson15a}, which however have been disputed~\cite{Pitkin,Desai}.

Here, we use median statistics technique (introduced by Gott et al~\cite{Gott} (hereafter G01) to combine results from all the latest measurements of $H_0$ and $G$  to  calculate the central values and associated  uncertainties.  We note that this technique was previously applied to about 553 measurements of $H_0$ as of 2012 by Chen and Ratra~\cite{Chen} (hereafter R11). Median statistics has also been used to estimate values of other cosmological parameters~\cite{Crandall}, in SN1a data analysis~\cite{Barreira}, mean matter density~\cite{Chen03}, and cosmological parameter estimation from SZ and X-ray observations~\cite{Sereno}.

The outline of this paper is as follows. In Sec.~\ref{sec:med}, we introduce the concept of median statistics, including calculation of confidence intervals and   and demonstrate its advantages over mean statistics using toy numerical experiments. We then apply the median statistics technique to measurements
of $H_0$ in Sec.~\ref{sec:H0} and $G$ in Sec.~\ref{sec:G}. We conclude in Sec.~\ref{sec:conclusions}.

\section{Median statistics}
\label{sec:med}
\par The central limit theorem~\cite{NR} says that under certain conditions, the arithmetic mean of a sufficiently large number of iterations of independent random variables, each with a well-defined (finite) expected value and finite variance, will approximately be normally distributed, regardless of the underlying parent distribution. If we consider each measurement ($X$) to be independent, we can expect the distribution of $X$ to be Gaussian.
\par One of the interesting features of  a Gaussian distribution is that it's mean and median are exactly the same. This opens up the possibility of using the median of a dataset for measurements when the mean is not adequate. Median statistics estimate can be obtained by minimizing sum of absolute deviation or L1 norm~\cite{NR}.
In the forthcoming sections we build upon this idea further using toy numerical simulations and finally use the \textbf{``Median Statistics"} technique to obtain best-fit values of  $H_0$ and $G$. For an accurate treatment of the median statistics on a given dataset, one should make the following assumptions:
\begin{description}
\item [\textbf{All the measurements are independent}]
This implies that the rank of the next observation is random and is equally likely to happen between any of the previous measurements.
\item [\textbf{There is no overall systematic error}]
If there exists an overall error, which affects all the  observations with the same magnitude as the statistical error, median statistics will lead to the wrong result (as we shall show at the end of \autoref{sec:se}). An important point to realize here is that even though  there cannot be an error affecting the whole data, it is possible to have different errors and sources of errors affecting mutually exclusive groups of data. \emph{We elaborate  on this further in} \autoref{sec:se}.
In case of the Gaussian distribution analogy discussed before, the bell curve would be shifted and hence, wouldn't be equal to the TRUE VALUE. Median statistics is immune to outliers, which otherwise affect our mean.
In Gott et al~\cite{Gott}, they have shown for a  Cauchy distribution how the robust median and the 95\% c.l.
are well-behaved unlike the mean and variance, which are plagued by outliers. They have calculated the empirical mean and c.l., which they compared with the median and it’s c.l. They note that c.l. range for mean is lot more than that for median. This shows the robustness of Median statistics. The mean of a dataset can be easily biased by adding/removing few extreme values, however the median remains insensitive.
\item [\textbf{The median is agnostic to  measurement errors}]
\emph{We shall elaborate upon  the ``watch'' analogy (originally introduced by Zeldovich)  described in  G01.} Let us consider  nine  watches for which the measurement of time from each watch is independent and its associated error is known $apriori$. The times shown by each watch  and associated uncertainties can be found  in  \autoref{tab:rw} with the true time equal to 1:00 PM. 
\begin{table}
\caption{Time read by nine watches along with the  uncertainties of each watch in minutes. We also provide a unique ``Build ID'' for each watch in the third column. These uncertainties are chosen in an ad-hoc fashion  with one outlier deliberately introduced.}

\label{tab:rw}
\begin{tabular}[t]{ l | c | r }
\hline
\textbf{Time} & \textbf{Uncertainty (in minutes)} & \textbf{Build ID} \\ \hline
11:50 AM & $\pm$ 25 & 2 \\ \hline
12:49 PM & $\pm$ 55 & 2 \\ \hline
12:52 PM & $\pm$ 4 & 3 \\ \hline
12:59 PM & $\pm$ 36 & 3 \\ \hline
01:02 PM & $\pm$ 7 & 1 \\ \hline
01:07 PM & $\pm$ 5 & 2 \\ \hline
01:10 PM & $\pm$ 70 & 1 \\ \hline
01:27 PM & $\pm$ 240 & 3 \\ \hline
02:35 PM & $\pm$ 3 & 1 \\ \hline
\end{tabular}
\end{table}
\par We find that the median time, 1:02 PM, turns out to be something close to the true value, viz. 1:00 PM. We need to employ different ways to calculate the uncertainty in median. Sufficient care has been taken to ensure that the time the watches read and the uncertainties they have are consistent. \emph{On an average, we expect watches to be inaccurate to within few  minutes. So, uncertainties are chosen in such a way that the  TRUE time lies in the interval around the watch time with uncertainty about the same as the half width.}

\subsection{Evaluating confidence levels}
\label{ss:ecl}
We briefly summarize the procedure for calculating the median confidence levels (c.l.), following G01 and R11, wherein more details can be found. The median of an array is defined as the $50\%$ percentile. Given $x \in $ array, then, there would be $i$ elements which are $\leq x$ and $(n-i)$ elements $ > x$ for some $i$, where $n$ is the length of the array.

Since we know that each given measurement is random and its rank completely arbitrary, we can
show that there are   ${n \choose i}$ ways to choose $i$ from $n$ measurements.
Furthermore, we assume that each measurement has equal probability of being $ \leq x$ or $ > x$,  or has a probability  equal to $\frac{1}{2}$.
Let $P_i$ be the probability that $i$ element to be the TRUE MEDIAN. $P_i$ is then given by:
\begin{equation}
P_i = \frac{1}{2^{N}}{n \choose i}
\end{equation}
\par Using this formula, we compute $P_i$ at every $i$ and define $C_j$ as sum of all $P_i$'s for $i$ ranging from $j:(N-j)$. To evaluate the 95\% confidence limit indices about the median, we choose the minimum value of  $r$ for which $C_r \geq 0.95$ where  $C_r \in (C_j:C_{N-j})$ for all values of  $j$ between 1 and $\frac{N}{2}$.

Using the above procedure, we find that the $95\%$ confidence limits ranges from 12:49 PM to 01:10 PM.
Our error is directly proportional to $1/\sqrt{N}$~\cite{Chen}. This is the same as what we would  expect from Gaussian mean statistics. This implies that our median statistics is able to obtain this value without positing anything about the error distribution.
\item[\textbf{Median Confidence limits}]
We understand from the  methodology discussed in the previous point, that the median confidence limit is a function of the sample size, $N$ and does not depend, in any way, on the sample distribution. In the limiting case, when $N = \lim_{n\to\infty}$, we find that the $100\%$ confidence limits index \emph{($r$ in the previous section)} is that of the median only.
This agrees with our intuition that larger the size of our dataset, smaller is the range of $95\%$ confidence intervals. To further drive home this point,  see \autoref{tab:rob}.
\begin{table*}[t]
\caption{\textbf{Robustness of Median statistics:} Here, we generate random samples from three separate distributions with different sizes and  means. Then, we compute the empirical mean and  median of each of the samples and tabulate the result. We notice that our estimate based on mean statistics is sensitive  to outliers, whereas our median is approximately equal to the true mean.} \label{tab:rob}
\begin{tabular}[b]{ l | c | c | c | c | c }
\hline
\textbf{Distribution} & \textbf{Sample size} & \textbf{Parameters} & \textbf{Calculated Mean} & \textbf{Median} & \textbf{95\% c.l. for Median} \\ \hline
\multirow{3}{*}{Gaussian} & $1000$ & $-10$ & $-9.997$ &$-9.993$ & $[-9.996, -9.991]$ \\ \cline{2-6}
& $2000$ & $10$ & $10.020$ &$10.029$ & $[10.027, 10.031]$ \\ \cline{2-6}
& $3000$ & $0$ & $0.0107$ &$0.0104$ & $[0.0092, 0.0106]$ \\ \hline
\multirow{3}{*}{Cauchy} & $1000$ & $-10$ & $-10.201$ &$-10.1384$ & $[-10.148, -10.1381]$ \\ \cline{2-6}
& $2000$ & $10$ & $8.513$ &$10.044$ & $[10.043, 10.045]$ \\ \cline{2-6}
& $3000$ & $0$ & $-2.234$ &$-0.007$ & $[-0.008, -0.006]$ \\ \hline
\multirow{3}{*}{Pareto with $\alpha \leq 1\ ,\ x_m = 1$} & $1000$ & $1/2$ & $15549.92$ &$3.663$ & $[3.660, 3.665]$ \\ \cline{2-6}
& $2000$ & $1/3$ & $607468177$ &$8.465$ & $[8.389, 8.494]$ \\ \cline{2-6}
& $3000$ & $1/4$ & $8.663e+12$ &$16.079$ & $[16.047, 16.096]$ \\ \hline
\end{tabular}
\end{table*}
\end{description}
\par An alternative to our approach of Median statistics would be Bayesian analysis~\cite{Parkinson,Trotta} (and references therein), wherein one uses Bayes' rule to determine the conditional probability of the correct observation \emph{(measurement)} given the other observations in hand. Although detailed comparison of the two methods is beyond the scope of this paper and depends on the data been analyzed, usually Bayesian parameter estimation is expected to be more accurate compared to  median estimates. In Bayesian analysis, one needs to posit a prior (based on previous measurements or expectations from theoretical models) used for the parameters or the model. Bayesian analysis is  also computationally expensive compared to the calculation of the median. The Bayesian approach  was first used  by Press~\cite{Press}  for $H_0$ values to marginalize over the bad measurements. Most recently,  $H_0$ was determined from Cepheid data by using Bayesian hyper-parameters~\cite{Cardona}.

\subsection{Effects of systematic errors}
\label{sec:se}
 \par One of the strongest assumptions we  make about the dataset is that it contains no overall egregious errors. This assumption does not preclude  us from saying that mutually exclusive groups may have  similar systematic errors affecting them. If one were to extend the nine  watches analogy to demonstrate this effect, one would have to provide \emph{extra information} about the watches. This \emph{extra information} is usually some way to classify the types of watches for example, eg. Model number, Batch code, Manufacturer name, etc. The only constraint the \emph{extra information} has to satisfy is to be able to classify each observation uniquely into mutually exclusive groups. The result can be summarized in  \autoref{tab:gr}.
\begin{table}
\caption{Here, we group all the watches based on their ``Build ID'' and calculate the group median. We then calculate the median of medians and show that this value is very close to the true value. We note that this simulated data has been  designed (by choice) so that the median of medians results in a better estimate of the true value.}
\label{tab:gr}
\begin{tabular}{ c | c | c | c }
\hline
Build ID & Group & Group Median & Median of Medians\\ \hline
\multirow{3}{*}{1} & 01:02 PM & \multirow{3}{*}{01:10 PM}  & \multirow{9}{*}{12:59 PM}  \\ \cline{2-2}
& 01:10 PM & & \\ \cline{2-2}
& 02:35 PM & & \\ \cline{2-2} \cline{1-3}
\multirow{3}{*}{2} & 11:50 AM & \multirow{3}{*}{12:49 PM} & \\ \cline{2-2}
& 12:49 PM & & \\ \cline{2-2}
& 01:07 PM & & \\ \cline{2-2} \cline{1-3}
\multirow{3}{*}{3} & 12:52 PM & \multirow{3}{*}{12:59 PM} & \\ \cline{2-2}
& 12:59 PM & & \\ \cline{2-2}
& 01:27 PM & & \\ \cline{2-2}
\hline
\end{tabular}
\end{table}
 \par Given any additional information about the nine watches, our estimate of the true median is more accurate than evaluating the median of the whole data, which gives a value of  1:02 PM. To summarize, we have taken into account the systematic error which affects the mutually exclusive groups encompassing our data.

\par We should point out one caveat related to the median estimate if the data contains systematic errors with a simple example. If  we add an offset of 30 minutes to the time shown by each watch in our dataset, our median of medians changes to 01:29 PM. Since the true value is  close to 01:00 PM  we can essentially get any value for the median of medians by adding/subtracting an arbitrary value to all our watches. A way to counter this is to hypothesize that there are NO overall systematic errors affecting our entire dataset.
 \par We now apply the median statistics method to measurements of Hubble's constant~$(H_0$) and Newton's Gravitational Constant~$(G)$.
\section{$H_0$ : Hubble's constant}
\label{sec:H0}
G01 introduced the notion of median statistics and carried out this analysis for $331$ published estimates of $H_0$. This was updated by  R11, using  553 measurements of  $H_0$ compiled by J. Huchra.
We replicate the analysis of R11 with updated measurements and present  median statistics estimates using $576$ values of $H_0$ (updated as of Sept. 2016) . We do not include any errors in our analysis. The full list of all $H_0$ measurements is uploaded on  \href{https://docs.google.com/spreadsheets/d/103HiCp2UFT4IgPQ6xJGHJXQs2LreAD0970Qj9In8tDM/edit?usp=sharing}{google docs}.

\par  The median value of all $H_0$ measurements along with 95\% c.l. without any grouping is given by $H_0=69.75\pm 5.25$ km/sec/Mpc.
The \emph{extra information} in the context of $H_0$ becomes the methodology employed to estimate $H_0$.
 \par R11  classified all measurements of $H_0$ into $18$ primary types and $5$ secondary types.
 Using this well-maintained list, one can calculate the median of each subgroup and also the median
 of all these medians. Furthermore, one can even calculate the 95\% c.l. as
 explained in \autoref{ss:ecl}. The \emph{primary types} grouping has been done based on the procedure employed to measure $H_0$ and is therefore  a good way to classify measurements. The \emph{secondary types} are more concerned with non-procedural factors involved in the measurement,
namely, determinations by one group (Sandage), indirect measurements based on the underlying  cosmological models, etc.
\par In \autoref{tab:t1}, we have grouped all measurements according to \emph{primary type}. And, in \autoref{tab:t3}, we do the same but group on the basis of \emph{secondary type}. Similar to R11, we also calculate the median values after excluding all measurements of \emph{primary} and \emph{secondary type}.
These are shown in ~\autoref{tab:t2} and ~\autoref{tab:t4} respectively.
The sub-group median values of both the primary and secondary types are shown in ~\autoref{tab:con}. The median estimate after grouping according to primary type is given by $H_0=72.5^{+2.5}_{-8}$ km/sec/Mpc. After grouping according to secondary type, we get  $H_0=68^{+4.5}_{-15.5}$ km/sec/Mpc.
 \begin{table*}[t]
  \centering
\caption{\textbf{$H_0$:Primary type grouping.} Here, we group the updated  list of $H_0$ measurements  on the basis of ``Primary type'' and perform median statistics analysis. ``Type'' is the class name. ``Number'' is the strength of the group. ``Median'' is the median of that group. ``95\% Confidence limits'' is the lower and upper confidence limits. The categories are the same as in R11.}
\label{tab:t1}
\begin{tabular}{l|c|c|c|c}
\hline
\textbf{Type} & \textbf{Number} & \textbf{Median} & \multicolumn{2}{c}{\textbf{95\% Confidence limits}} \\ \hline
Global Summary & 118 & 70.0 & 69.0 & 71.0 \\ \hline
Type Ia Supernovae & 97 & 64.0 & 60.0 & 66.0 \\ \hline
Other & 85 & 68.0 & 60.0 & 71.0 \\ \hline
Lens & 84 & 64.5 & 62.0 & 69.0 \\ \hline
Sunyaev-Zeldovich & 46 & 60.5 & 57.0 & 66.0 \\ \hline
Baryonic Tully-Fisher & 23 & 60.0 & 56.0 & 71.0 \\ \hline
Infrared Tully-Fisher & 19 & 82.0 & 65.0 & 90.0 \\ \hline
Fluctuations & 18 & 75.0 & 71.0 & 81.0 \\ \hline
Tully-Fisher & 18 & 72.5 & 68.0 & 74.0 \\ \hline
CMB fit & 16 & 69.5 & 58.0 & 71.0 \\ \hline
Globular Cluster Luminosity functions & 14 & 76.5 & 65.0 & 80.0 \\ \hline
$D_n-\sigma$/Fund plane & 10 & 75.0 & 67.0 & 78.0 \\ \hline
Inverse Tully-Fisher & 9 & 74.0 & 69.0 & 77.0 \\ \hline
Type II Supernovae & 8 & 59.5 & 52.0 & 76.0 \\ \hline
Planetary Nebula Luminosity Functions & 6 & 85.0 & 77.0 & 87.0 \\ \hline
Novae & 4 & 77.0 & - & - \\ \hline
Red Giants & 1 & 74.0 & - & - \\ \hline
\end{tabular}

\caption{\textbf{$H_0$:Complement of Primary type grouping.}
Here, we form a set of measurements which do not belong to a specific ``Primary Type''
and perform median statistics analysis. ``Type'' is the class name.
``Number'' is the strength of the group. ``Median'' is the median of that group.
95\% Confidence limits correspond to the lower and upper confidence intervals.}
\label{tab:t2}
\begin{tabular}{l|c|c|c|c}
\hline
\textbf{Type} & \multicolumn{1}{l}{\textbf{Number}} \vline & \multicolumn{1}{l}{\textbf{Median}} \vline & \multicolumn{2}{l}{\textbf{95\% Confidence limits}} \\ \hline
Red Giants & 574 & 68.95 & 68.11 & 69.0 \\ \hline
Novae & 571 & 68.9 & 68.0 & 69.0 \\ \hline
Planetary Nebula Luminosity functions & 569 & 68.0 & 68.0 & 68.11 \\ \hline
Type II Supernovae & 567 & 69.0 & 68.9 & 69.0 \\ \hline
Inverse Tully-Fisher & 566 & 68.0 & 68.0 & 68.0 \\ \hline
$D_n-\sigma$/Fund plane & 565 & 68.11 & 68.0 & 68.9 \\ \hline
Globular Cluster Luminosity Functions & 561 & 68.0 & 68.0 & 68.11 \\ \hline
CMB fit & 559 & 68.9 & 68.0 & 69.0 \\ \hline
Fluctuations & 557 & 68.0 & 68.0 & 68.0 \\ \hline
Tully-Fisher & 557 & 68.0 & 68.0 & 68.0 \\ \hline
Infrared Tully-Fisher & 556 & 68.0 & 68.0 & 68.0 \\ \hline
Baryonic Tully-Fisher & 552 & 69.0 & 69.0 & 69.0 \\ \hline
Sunyaev-Zeldovich & 529 & 69.0 & 69.0 & 69.0 \\ \hline
Lens & 491 & 69.0 & 69.0 & 69.0 \\ \hline
Other & 490 & 69.0 & 69.0 & 69.0 \\ \hline
Type Ia Supernovae & 478 & 69.0 & 69.0 & 69.0 \\ \hline
Global Summary & 458 & 68.0 & 67.0 & 68.0 \\ \hline
\end{tabular}
\caption{\textbf{$H_0$:Secondary type grouping.}
Here, we group on the basis of ``Secondary type'' and perform median statistics analysis.
``Type'' is the class name. ``Number'' is the strength of the group.
``Median'' is the median of that group. ``95\% Confidence limits'' is the lower and upper confidence limits.}
\label{tab:t3}
\begin{tabular}{l|c|c|c|c}
\hline
\textbf{Type} & \multicolumn{1}{l}{\textbf{Number}} \vline & \multicolumn{1}{l}{\textbf{Median}} \vline & \multicolumn{2}{l}{\textbf{95\% Confidence limits}} \\ \hline
No second type & 329 & 69.0 & 69.0 & 69.0 \\ \hline
Cosmology dependent & 84 & 68.0 & 67.0 & 68.0 \\ \hline
Sandage and/or Tammann & 71 & 55.0 & 55.0 & 56.0 \\ \hline
Key Project or Key Project team Member & 62 & 72.5 & 72.0 & 73.0 \\ \hline
deVaucouleurs or van den Bergh & 21 & 95.0 & 89.0 & 95.0 \\ \hline
results presented at Irvine Conf & 5 & 65.0 & - & - \\ \hline
Theory with assumed Omega & 4 & 52.5 & - & - \\ \hline
\end{tabular}
\end{table*}
\begin{table*}
\caption{\textbf{$H_0$:Complement set of Secondary type grouping.}
Here, we tabulate a set of measurements which do not belong to a specific ``Secondary Type''
and perform median statistics analysis. ``Type'' is the class name.
``Number'' is the strength of the group. ``Median'' is the median of that group. `
`95\% Confidence limits'' is the lower and upper confidence limits.}
\label{tab:t4}
\begin{tabular}{l|c|c|c|c}
\hline
\textbf{Type} & \multicolumn{1}{l}{\textbf{Number}} \vline & \multicolumn{1}{l}{\textbf{Median}} \vline & \multicolumn{2}{l}{\textbf{95\% Confidence Limits}} \\ \hline
Theory with assumed Omega & 571 & 69.0 & 68.9 & 69.0 \\ \hline
results presented at Irvine Conf & 570 & 69.0 & 69.0 & 69.0 \\ \hline
deVaucouleurs or van den Bergh & 554 & 68.0 & 68.0 & 68.0 \\ \hline
Key Project or Key Project team Member & 513 & 67.0 & 67.0 & 67.0 \\ \hline
Sandage and/or Tammann & 504 & 70.0 & 70.0 & 70.0 \\ \hline
Cosmology dependent & 492 & 69.0 & 69.0 & 69.0 \\ \hline
No second type & 246 & 67.0 & 67.0 & 67.0 \\ \hline
\end{tabular}
\end{table*}
\section{$G$ : Newton's Gravitational Constant}
\label{sec:G}
\par We now apply the median statistics technique to do similar analysis of  $G$. For this purpose, we use the tabulated measurements of $G$ from Schlamminger et al~\cite{Schlamminger}, who have a compiled a list of all $G$ measurements since 1980. A review of all previous measurements (starting from the first one by Cavendish in 1798) and associated controversies are reviewed in Refs.~\cite{Quinn,Mohr,Gilles}. The global median value of all measurements is given by $G=(6.674252^{+0.003655}_{-0.002342}) \times 10^{-11} \mathrm{N m^2 kg^{-2}}$, where the error bars correspond to 95\% c.l.
We now group the measurements, similar to what was done for $H_0$ according to how the measurements were made.
For our grouping criterion, we have considered the   \emph{device} used for measurement and the \emph{mode} in which the \emph{device} was used. In case of \emph{torsion balance}, the same setup can be used for different procedures. Similar to $H_0$, we catagorize the grouping done according to \emph{mode} and \emph{device} as \emph{primary} and \emph{secondary} respectively.

A tabular summary of  the median of  $G$ measurements after collating the observations into various groups can be found in \autoref{tab:g1} and \autoref{tab:g3}. \autoref{tab:g1} shows the results  when classification of all $G$ measurements is done only by \emph{device}. Similar grouping and tabulation are done for \emph{mode} in \autoref{tab:g3}. The median values after excluding all measurements of \emph{device} and \emph{mode} type are shown in \autoref{tab:g2} and \autoref{tab:g4} respectively. The sub-group median values of both the primary  and secondary types are shown in ~\autoref{tab:con}.  The median estimate after grouping according to primary type is given by $G=6.674702\times 10^{-11}~\mathrm{N m^2/kg^2}$, and after grouping by secondary type is equal to  $6.673765 \times 10^{-11}~\mathrm{N m^2/kg^2}$. Unlike $H_0$, we could not obtain 95\% c.l. intervals for the subgroup medians, since the total number of measurements were quite small. So we only calculate 68\% c.l. uncertainties for the sub-group medians for $G$, which can be found in ~\autoref{tab:con}.

\begin{table*}[t]
  \centering
\caption{\textbf{$G$:Device grouping-} We have grouped all measurements by ``Device'' and computed the group-wise median and 95\% confidence interval. ``Number'' denotes the strength of each group. \emph{Please note that Median and 95\% confidence limits have a multiplication factor of $\times 10^{-11}$ and units are in $\mathrm{N m^{2} kg^{-2}}$}}
\label{tab:g1}
\begin{tabular}{l|c|c|c|c}
\hline
\textbf{Device}  & \multicolumn{1}{l}{\textbf{Median x$10^{-11}$}} \vline & \multicolumn{1}{l}{\textbf{Number}} \vline & \multicolumn{2}{l}{\textbf{95\% Confidence interval x$10^{-11}$}} \\ \hline
Torsion Balance & 6.674255  & 17 & 6.67346  & 6.67553  \\ \hline
Two Pendulums & 6.67328 & 2 & - & - \\ \hline
Atom Interferometer & 6.67191  & 1 & - & - \\ \hline
Beam Balance & 6.67425 & 1 & - & - \\ \hline
\end{tabular}

\caption{\textbf{$G$:Complement set of Device Grouping-} We have considered all the measurements which do not belong to a ``device'' and calculated group-wise median along with 95\% confidence limits. ``Number'' denotes the strength of each group. \emph{Please note that Median and 95\% confidence limits have a multiplication factor of $\times 10^{-11}$ and units are in $\mathrm{N m^{2} kg^{-2}}$}}
\label{tab:g2}
\begin{tabular}{l|c|c|c|c}
\hline
\textbf{Device}  & \multicolumn{1}{l}{\textbf{Median x$10^{-11}$}} \vline & \multicolumn{1}{l}{\textbf{Number}} \vline & \multicolumn{2}{l}{\textbf{95\% Confidence interval $ \times 10^{-11}$}} \\ \hline
Atom Interferometer & 6.674235  & 20 & 6.67352  & 6.67455  \\ \hline
Beam Balance & 6.67415  & 20 & 6.67346  & 6.67455  \\ \hline
Two Pendulums & 6.67425  & 19 & 6.673460  & 6.67515  \\ \hline
Torsion Balance & 6.67328 & 4 & - & - \\ \hline
\end{tabular}

\caption{\textbf{$G$:Mode grouping-} We have grouped by ``Mode'' and computed group-wise median and 95\% confidence limits. ``Number'' denotes the strength of each group. \emph{Please note that Median and 95\% confidence limits have a multiplication factor of $\times 10^{-11}$ and units are in $\mathrm{N m^{2} kg^{-2}}$}}
\label{tab:g3}
\begin{tabular}{l|c|c|c|c}
\hline
\textbf{Mode}  & \multicolumn{1}{l}{\textbf{Median x$10^{-11}$}} \vline & \multicolumn{1}{l}{\textbf{Number}} \vline & \multicolumn{2}{l}{\textbf{95\% Confidence interval $\times 10^{-11}$}} \\ \hline
time of swing & 6.67352 & 9 & 6.6729 & 6.674  \\ \hline
No Mode given & 6.67328 & 4 & - & - \\ \hline
electrostatic servo & 6.67515 & 3 & - & - \\ \hline
Cavendish & 6.675755  & 2 & - & - \\ \hline
Cavendish and servo & 6.675565 & 2 & - & - \\ \hline
acceleration servo & 6.674255  & 1 & - & - \\ \hline
\end{tabular}

\caption{\textbf{$G$:Complement set of Mode grouping-} As done before, we have analyzed all the measurements which do not belong to a ``mode'' and computed groupwise median and 95\% confidence limits. ``Number'' denotes the strength of each group. \emph{Please note that Median and 95\% confidence limits have a multiplication factor of $10^{-11}$ and units are in $\mathrm{N m^{2} kg^{-2}}$}}
\label{tab:g4}
\begin{tabular}{l|c|c|c|c}
\hline
\textbf{Mode}  & \multicolumn{1}{l}{\textbf{Median x$10^{-11}$}} \vline & \multicolumn{1}{l}{\textbf{Number}} \vline & \multicolumn{2}{l}{\textbf{95\% Confidence interval $\times 10^{-11}$}} \\ \hline
acceleration servo & 6.67415 & 20 & 6.67346  & 6.67455 \\ \hline
Cavendish & 6.67408   & 19 & 6.6729 & 6.67435 \\ \hline
Cavendish and servo & 6.67408  & 19 & 6.6729 & 6.67435 \\ \hline
electrostatic servo & 6.67415  & 18 & 6.6729 & 6.67435 \\ \hline
No mode given & 6.674255 & 17 & 6.67352  & 6.67515 \\ \hline
time of swing & 6.6747025 & 12 & 6.67387 & 6.67554 \\ \hline
\end{tabular}
\end{table*}

\section{Conclusions}
\label{sec:conclusions}
In this article, we have described the  usage of  median statistics technique used to obtain the central values of parameters along with associated 95\% c.l. uncertainties. We have demonstrated its robustness  over mean statistics using numerical experiments. We have extended previous median statistics measurements by R11 to the  full list of 576 Hubble Constant ($H_0$) measurements (updated as of Sept. 2016).  We grouped  all measurements according to primary and secondary categories (using the same classification as R11) and estimated the median value in each category. We then calculated the median of all these sub-group medians in both the categories. We find that for the primary type of measurements, the sub-group median estimate is given by  $H_0 = 72.5 \ \mathrm{km s^{-1} Mpc^{-1}}$ with $95\%$ confidence limits between $ 64.5$ and $75.0 $ km/s/Mpc. We then carried out the same exercise for all measurements of Newton's Gravitational Constant ($G$), tabulated since 1980.  Here, the grouping was done on basis of the device used for measurements and the mode of measurement (for any given device). We find that the sub-group median (after splitting according to  mode of measurement) is given by  $G = 6.674702 \times 10^{-11}~\mathrm{N m^2 kg^{-2}}$ and $68\%$ confidence limits are $[6.671910, 6.675565] \times 10^{-11} \mathrm{N m^2 kg^{-2}}$. The summary statistics of $H_0$ and $G$ are tabulated in  ~\autoref{tab:con}.
\begin{table*}[b]
\caption{This table summarizes the outcome of this paper. Here, groups correspond to the ``group by''\emph{Primary and secondary type} in case of $H_0$ and, \emph{Device and Mode} in case of $G$. Since, each constant has one ``group by'', there are total of two cases for each constant. For $H_0$, the first two values in ``Table Wise Median'' column correspond to \textbf{Primary type},  \textbf{Secondary type.} In case of $G$, the two values correspond to \textbf{Mode}, \textbf{Device}. We are specifying 68\% c.l. only for \emph{Mode} ``group'' as the strength of the group is too less to compute 95\% c.l. For the remaining case, c.l. corresponds to 95\% c.l. In the \emph{Median} and \emph{95\% c.l}, we take global median and 95\% c.l. using the two tables we computed for each Fundamental Constant. \emph{Please note that for $G$, the values in all the rows corresponding to it have a multiplication factor of $\times 10^{-11}$.}}
\label{tab:con}
\begin{tabular}{c|c|c|c|c|c|c}
\hline
\multicolumn{1}{l}{\textbf{Constant}} \vline & \textbf{Group Median} & \multicolumn{2}{l}{\textbf{c.l.}} \vline & \textbf{Global Median} & \multicolumn{2}{l}{\textbf{95\% c.l.}} \\ \hline
\multirow{2}{*}{$H_0$} & 72.5 & 64.5 & 75.0 & \multirow{2}{*}{$ 69.75$ km/s/Mpc} & 64.5 & 74.0\\ \cline{2-4}
 & 68 & 52.5 & 72.5 & &  \\ \hline
\multirow{2}{*}{$G$} & 6.674702 & 6.67328  & 6.675565  & \multirow{2}{*}{6.674252 $ \mathrm{N m^2 kg^{-2}}$} & 6.67191 & 6.675565 \\ \cline{2-4}
 & 6.673765 & 6.67191 & 6.67425 & & \\ \hline 
\end{tabular}
\end{table*}
\acknowledgement{{\bf Acknowledgements:} We would like to thank  Gang Chen for sharing with us the data used in R11.}

\bibliographystyle{spphys}       
\bibliography{Med_stat}   
\end{document}